\documentclass[sn-mathphys]{sn-jnl}% Math and Physical Sciences Reference Style
\jyear{2021}%

\theoremstyle{thmstyleone}%
\theoremstyle{thmstyletwo}%

\theoremstyle{thmstylethree}%

\usepackage{graphicx}
\usepackage[center]{caption}
\usepackage{mathptmx}      % use Times fonts if available on your TeX system
\usepackage{amsmath,amssymb}
\begin{document}

\title[Comparison of quantum state protection against decoherence, a survey]{Comparison of quantum state protection against decoherence via weak measurement, a survey}


\author*[1]{\fnm{Sajede} \sur{Harraz}}\email{sajede@ustc.edu.cn}

\author[1]{\fnm{Shuang} \sur{Cong}}\email{scong@ustc.edu.cn}

\author[2]{\fnm{Juan J.} \sur{Nieto}}\email{juanjose.nieto.roig@usc.es}

\affil*[1]{\orgdiv{Department of Automation}, \orgname{University of Science and Technology of China}, \orgaddress, \city{Hefei}, \postcode{230027}, \country{China}}
\affil[2]{\orgdiv{Instituto de Matemáticas, CITMAga}, \orgname{Universidade de Santiago de Compostela}, \orgaddress{\city{Santiago de Compostela}, \postcode{15782}, \country{Spain}}}


\abstract{One of the crucial tasks in quantum systems is to reduce the effects of decoherence due to the unavoidable interactions between a system and its environment. Many protection schemes have been proposed recently, among them the weak measurement quantum measurement reversal (WMQMR), weak measurement-based quantum feedback control (QFBC) and quantum feed-forward control (QFFC) are reviewed in this paper. By considering weak measurement, the aim is to find a balance between information gain and disturbance of the system caused by the measurement. We classify different types of measurement and give the definition of noise sources and their effects on the state of the system. Finally, we compare and analyze the performance of the discussed protection schemes for different noise sources by numerical simulations.}




\maketitle

\section{Introduction}\label{sec1}

The control of a system is trying to reach some objective. The objective can be reaching a special state at a given time or making the system behave in a certain way. In quantum state protection, the control strategies use some rules to protect the state from environmental noise \cite{lab1,lab2,lab2a,lab2b,lab2c,lab2d}.  In realistic situations, the system is not isolated and affected by the environment inevitably, which disturbs the coherence of the system \cite{lab3,lab3a,lab3b,lab3c}. Therefore it is fundamentally important to study the protection of the quantum state from environmental noise.  In classical systems, one can protect the state from noise by taking complete record of the state before noise. However, in quantum systems one can not gain information about the system without disturbing it \cite{lab4,lab4a,lab4b,lab4c}. An essential feature of quantum measurements is the disturbance to the system caused by measurement which is the so-called quantum back-action \cite{lab5,lab6}. The traditional von Neumann orthogonal projection measurement is irreversible since it collapses the state of the system completely to its eigenstate. While weak measurement (WM) can decrease the disturbance of the system by weakening the interaction responsible for the measurement \cite{lab5a,lab6a,lab6b,lab6c,lab6d}. There is a tradeoff between information gain and disturbance of the system in quantum WM. The weaker measurement gives less information about the system and disturbs it less \cite{lab7a,lab7b}. By using quantum WM, quantum control strategies are available that exceed schemes based on classical concepts. Hence, WM has been widely used in protection of quantum states against decoherence \cite{lab8a,lab9,lab10,lab11,lab12}. 

There are many types of decoherence, while phase damping (PD) and amplitude damping (AD) are typical decoherence mechanisms, which occurs in many practical quantum systems and has aroused wide concern \cite{lab13,lab14,lab15}. In the PD channel, the decoherence happens as some random phase shifts of the system which is caused by its interaction with the environment. In the AD channel, the initial state will gradually decay to a steady state related to the temperature of the environment.

A number of methods have been proposed to protect the state from decoherence, such as: the so-called decoherence-free subspace \cite{lab16,lab17,lab18}; however, it requires the interaction Hamiltonian to have an appropriate symmetry, which might not always be present. Another method doing quantum state protection is referred to as quantum error correction \cite{lab19,lab20,lab21}. However, it needs redundant abstract information to encode the physical system to help states get through the noise without being affected which needs more resource cost.

Another interesting method is weak measurement and quantum measurement reversal (WMQMR) which can protect the state of the system without additional resources \cite{lab7,lab22,lab23,lab23a}. Before the decoherence process, by applying the WM the state is transferred to a more robust state to resist the decoherence. After the decoherence, the quantum measurement reversal (QMR) is applied to reverse back the state to its initial one. These schemes are probabilistic due to the probability rate of the WM. Moreover, there are two major techniques to protect the quantum state against decoherence: quantum feedback control (QFBC) \cite{lab24,lab25} and quantum feed-forward control (QFFC) \cite{lab26,lab27}. In QFBC, after the state passes through the noise channel a measurement is applied to find some information about the effects of the noise and the state of the system. Then, according to the result of the measurement a control operation is applied to recover the state. While in QFFC, firstly the quantum state is transferred to a more robust state to resist the decoherence. After the decoherence, the state is brought back to the original state by reversal operations. 

In most papers the QFBC is used to overcome the effects of PD noise while QFFC and WMQMR are applied to protect the state from AD. Here the question arises as to what would be the effects of each protection scheme in occurrence of different noise sources. To answer this question, we compare the performance of QFFC and QFBC in presence of AD and PD by numerical simulations. To do the comparison, we assume that the initial state and noise are fully characteristic and known.

The state protection schemes have applications in entanglement protection to improve the entanglement in presence of noise. Quantum entanglement is an essential resource for quantum information processing including quantum computation \cite{lab28,lab29} and quantum communication \cite{lab30,lab31}. However, in realistic implementations, noise is affecting the entangled state due to irreversible interactions with the environment, and in some cases, leads to entanglement sudden death (ESD) in which entanglement is completely lost before its subsystem is fully decohered \cite{lab32,lab33}. The state protection schemes can protect the entanglement as well and avoid ESD. 

Another application of protection schemes is in discrimination and protection of nonorthogonal states. When a system is prepared in one of several nonorthogonal states, no measurement can determine which preparation occurred with certainty \cite{lab34}. Hence, protecting nonorthogonal states from noise is a big challenge. The nonorthogonal qubit states are particularly important in the well-known B92 quantum key distribution protocol \cite{lab35,lab36,lab37}.  

This paper is organized as follows. In Sec. \ref{sec2}, we present different definitions of noise sources and the measurement and control operators, which are used in quantum state protection schemes. Sec. \ref{sec3}  is the review of the WMQMR scheme, and Sec. \ref{sec4} is the review of the protection schemes based on QFBC. In Sec. \ref{sec5} we review the QFFC protection schemes. Finally, the comparison between QFBC and QFFC protection schemes is given in Sec. \ref{sec6} followed by the conclusion in Sec. \ref{sec7}.

\section{Mathematical concepts}\label{sec2}

In this section, we give different definitions of two most important noise sources: AD and PD. Then, we present the measurement and control operators, which are used in quantum state protection schemes against decoherence. The operators may be named differently in different papers, but the definitions are all the same. Therefore, we will frequently use these defined operators in the following sections.

\subsection{ Quantum phase damping and amplitude damping}

The unwanted interactions with the environment in real quantum systems show up as noise. To protect the state of the quantum system against environmental noise sources, first we need to know their characteristics. In this subsection, we present the definition and effects of two typical noise sources: PD and AD. 

The PD noise can be defined by a quantum operation that acts on a single-qubit density matrix $\rho $ as:

\begin{equation}\label{eq1} 
\varepsilon _{r} (\rho )=r(Z\rho Z)+(1-r)\rho  
\end{equation}
where $Z=\left(\begin{array}{cc} {1} & {0} \\ {0} & {-1} \end{array}\right)$is the Pauli operator, and $r$ the damping probability that is known and without loss of generality in the range $0\le r\le 0.5$ . According to Eq. \eqref{eq1}, the PD noise leaves the system unchanged with probability $(1-r)$ and applies a phase flip $Z$ to the system with probability $r$. 

The PD noise can also be defined as a rotation with quantum operation:

\begin{equation}\label{eq2} 
\varepsilon _{\lambda } (\rho )=\sin ^{2} ({\lambda \mathord{\left/ {\vphantom {\lambda  2}} \right. \kern-\nulldelimiterspace} 2} )(Z\rho Z)+\cos ^{2} ({\lambda \mathord{\left/ {\vphantom {\lambda  2}} \right. \kern-\nulldelimiterspace} 2} )\rho  
\end{equation}
where $\lambda$ is the kicking angle applied by PD noise. Hence, the PD noise is considered as ``kicking'' the state of the qubit in one direction or the other $\pm \lambda $ on the Bloch sphere. This quantum operation is equivalent to Eq. \eqref{eq1} with $r=\sin ^{2} ({\lambda \mathord{\left/ {\vphantom {\lambda  2}} \right. \kern-\nulldelimiterspace} 2} )$. 

 The PD noise can also be written as a map:
\begin{equation}\label{eq3} 
\rho \to \varepsilon _{PD} \left(\rho \right)=A_{0} \rho A_{0}^{\dag } +A_{1} \rho A_{1}^{\dag }  
\end{equation}
with $A_{0} $ and $A_{1} $, the Kraus operators of the PD noise given by
\begin{equation}\label{eq4} 
A_{0} =\left(\begin{array}{cc} {1} & {0} \\ {0} & {\sqrt{1-r} } \end{array}\right)\, ,\, A_{1} =\left(\begin{array}{cc} {0} & {0} \\ {0} & {\sqrt{r} } \end{array}\right)\,  
\end{equation}
 
 To better understand the effect of PD noise on the state of the system, consider the state of a two-level quantum system in density matrix form as:
\begin{equation}\label{eq5} 
\rho =\left[ \begin{array}{cc}
{\rho }_{11} & {\rho }_{12} \\ 
{\rho }_{21} & {\rho }_{22} \end{array}
\right]=\frac{1}{2}\left[ \begin{array}{cc}
1+z & x+i*y \\ 
x-i*y & 1-z \end{array}
\right] 
\end{equation}
in which {$x, \ y,\ z$} are the Cartesian coordinates of a qubit. The relation between Cartesian coordinates and the elements of the state density matrix are: 
\begin{equation}\label{eq6} 
x=2*real\left({\rho }_{12}\right);y=2*imag\left({\rho }_{12}\right);z=2*{\rho }_{11}-1;
\end{equation}

For the general single qubit state $\rho $ in Eq. \eqref{eq5}, the state of the system after being affected by PD according to Eq. \eqref{eq3} can be described as \cite{lab38}:
\begin{equation}\label{eq7} 
{\rho }_e=A_0\ \rho {\ A}^{\dagger }_0+A_1\ \rho {\ A}^{\dagger }_1=\left[ \begin{array}{cc}
{\rho }_{11} & \sqrt{1-r}*\ {\rho }_{12} \\ 
\sqrt{1-r}*\ {\rho }_{21} & {\rho }_{22} \end{array}
\right]
\end{equation}

As one can see from Eq. \eqref{eq6} and Eq. \eqref{eq7}, PD changes the value of $x$ and $y$  components of the quantum state but leaves the value of $z$ component unchanged. Hence, the effect of the phase damping is to spoil the off-diagonal elements of the density matrix, or to shorten the $x$ and $y$ component of the Bloch vector for any state. Therefore, the effect of PD noise on the Bloch vector is to shorten the length of the Bloch vector (making the state less pure), and change the angle between the Bloch vector and one of the planes of the Bloch sphere (depending on the initial state).  

Another important type of decoherence which is related to many practical qubit systems is AD, which can happen to a photon qubit in a leaky cavity, or atomic qubit subjected to spontaneous decay, or a superconduction qubit with zero-temperature energy relaxation \cite{lab39}. Single-qubit AD  can be mathematically described by the following mappings \cite{lab13,lab40}:
\begin{equation}\label{eq8} 
\begin{array}{l} {{\lvert 0 \rangle} _{S} {\lvert 0 \rangle} _{E} \to {\lvert 0 \rangle} _{S} {\lvert 0 \rangle} _{E} } \\ {{\lvert 1 \rangle} _{S} {\lvert 0 \rangle} _{E} \to \sqrt{1-r} {\lvert 1 \rangle} _{S} {\lvert 0 \rangle} _{E} +\sqrt{r} {\lvert 0 \rangle} _{S} {\lvert 1 \rangle} _{E} } \end{array} 
\end{equation}
where $r\in [0,1]$ is the possibility of decaying the excited state, and  $S (E)$ denotes the system (environment). Within the Weisskopf--Wigner approximation, the probability of finding the atom in the excited state decreases exponentially with time, and we have $\sqrt{1-r} =e^{-\Gamma t} $ with $\Gamma $ being the spontaneous decay rate of the atom. 

If we only consider the state evolution of the atom, the amplitude damping process is described by a map from an input state $\rho _{in} $ to an output state $\rho ^{AD} $ \cite{lab41}:
\begin{equation}\label{eq9} 
\rho _{in} \to \rho ^{AD} =A_{0} \rho _{in} A_{0}^{\dag } +A_{1} \rho _{in} A_{1}^{\dag }  
\end{equation}
where the AD Kraus operators are given by
\begin{equation}\label{eq10} 
A_{0} =\left(\begin{array}{cc} {1} & {0} \\ {0} & {\sqrt{1-r} } \end{array}\right)\, ,\, A_{1} =\left(\begin{array}{cc} {0} & {\sqrt{r} } \\ {0} & {0} \end{array}\right)\,  
\end{equation}

To consider the AD, there is the mathematical trick of unraveling \cite{lab23}. That is, the qubit trajectories through the AD are considered as two parts, the jump one $(A_{1} )$ and no jump one$(A_{0} )$. The state of the system after each part is given as 
\begin{equation}\label{eq11} 
\rho _{no\, jump} =\left(\begin{array}{cc} {\rho _{11} } & {\rho _{12} \sqrt{1-r} } \\ {\rho _{21} \sqrt{1-r} } & {\rho _{22} (1-r)} \end{array}\right)\, ,\, \, \rho _{jump} =\left(\begin{array}{cc} {r\rho _{22} } & {0} \\ {0} & {0} \end{array}\right) 
\end{equation}

As Eq. \eqref{eq11} depicted, the state of the system will jump to state $\rho _{jump} \approx {\lvert 0 \rangle} {\langle 0 \rvert} $ with probability $r\rho _{22} $ and no jump to state $\, \rho _{no\, jump} $ with probability $(1-r)$.   

\subsection{ The measurement and control operators  }

To gain information about the state of the system one needs to use quantum measurement. In most protection schemes a family of positive operator-valued measurement (POVM) consists of two operators given by $E_{m} =M_{m}^{\dag } M_{m} $is used.  The measurement operators $M_{m} $ among different axes of the Bloch sphere are as follow:  
\begin{equation}\label{eq12} 
\begin{aligned}
M_{x\pm } \left(\theta \right)&=\cos \left(\theta /2\right){\lvert \pm  \rangle} {\langle \pm  \rvert} +\sin \left(\theta /2\right){\lvert \mp  \rangle} {\langle \mp  \rvert} \\
M_{y\pm } \left(\theta \right)&=\cos \left(\theta /2\right){\lvert \pm i \rangle} {\langle \pm i \rvert} +\sin \left(\theta /2\right){\lvert \mp i \rangle} {\langle \mp i \rvert}  \\
M_{z+} \left(\theta \right)&=\cos \left(\theta /2\right){\lvert 0 \rangle} {\langle 0 \rvert} +\sin \left(\theta /2\right){\lvert 1 \rangle} {\langle 1 \rvert} \\
M_{z-} \left(\theta \right)&=\sin \left(\theta /2\right){\lvert 1 \rangle} {\langle 1 \rvert} +\cos \left(\theta /2\right){\lvert 0 \rangle} {\langle 0 \rvert} 
\end{aligned}
\end{equation}
where$\theta \in \left[0,\pi /2\right]$, ${\lvert \pm  \rangle} =\frac{1}{\sqrt{2} } \left({\lvert 0 \rangle} \pm {\lvert 1 \rangle} \right)$ and ${\lvert \pm i \rangle} =\frac{1}{\sqrt{2} } \left({\lvert 0 \rangle} \pm i{\lvert 1 \rangle} \right)$. $M_{x\pm } \left(\theta \right)$ are measurement operators along the $x$-axis, $M_{y\pm } \left(\theta \right)$ along the $y$-axis and $M_{z\pm } \left(\theta \right)$ along the $z$-axis. $\theta $. When $\theta ={\raise0.7ex\hbox{$ \pi  $}\!\mathord{\left/ {\vphantom {\pi  2}} \right. \kern-\nulldelimiterspace}\!\lower0.7ex\hbox{$ 2 $}} $, the strength of the measurement is zero and we call it as : ''no measurement'' (NM) scheme. For $\theta =0$, the strength of the measurement is maximum and is called ``projective measurement'' (PM), in a way that the state of the system is projected onto its eigenstates. When $0<\theta <{\raise0.7ex\hbox{$ \pi  $}\!\mathord{\left/ {\vphantom {\pi  2}} \right. \kern-\nulldelimiterspace}\!\lower0.7ex\hbox{$ 2 $}} $ , we call it the ``weak measurement'' (WM) scheme. Hence, the strength of the measurement is adjustable for the trade-off between the information gain and disturbance of the system because of measurement.

The experimental realization of the WM is discussed theoretically in \cite{lab42} where the WM can be realized by coupling the system to a meter and performing the usual projective measurements on the meter only. Also, the experimental implementation is realized in a photonic architecture \cite{lab26}. A measurement of this kind can also be realized in nuclear magnetic resonance by means of coupling the spin under consideration to one of its neighbors \cite{lab43}.

 After gaining information about the state of the system, one needs to apply the control operations. Most protection schemes use rotation as control operation. The unitary rotations along different axes of the Bloch sphere are defined as
\begin{equation}\label{eq13} 
\begin{aligned}
R_{x} \left(\pm \eta \right)&=\left[\begin{array}{cc} {\cos \left(\eta /2\right)} & {\mp i*\sin \left(\eta /2\right)} \\ {\pm {\rm i*}\sin \left(\eta /2\right)} & {\cos \left(\eta /2\right)} \end{array}\right]\\
R_{y} \left(\pm \eta \right)&=\left[\begin{array}{cc} {\cos \left(\eta /2\right)} & {\mp \sin \left(\eta /2\right)} \\ {\pm \sin \left(\eta /2\right)} & {\cos \left(\eta /2\right)} \end{array}\right] \\
R_{z} \left(\pm \eta \right)&=e^{i\eta {Z\mathord{\left/ {\vphantom {Z 2}} \right. \kern-\nulldelimiterspace} 2} } =\left[\begin{array}{cc} {e^{\pm \; i\; \frac{\eta }{2} } } & {0} \\ {0} & {e^{\mp i\; \frac{\eta }{2} } } \end{array}\right]
\end{aligned}
\end{equation}
where $\eta $ is the rotation angle in the range $0\le \eta \le {\pi \mathord{\left/ {\vphantom {\pi  2}} \right. \kern-\nulldelimiterspace} 2} $, $R_{x} \left(\pm \eta \right)$ represents the rotation along the $x$-axis, $R_{y} \left(\pm \eta \right)$ along the $y$-axis and $R_{z} \left(\pm \eta \right)$ along the $z$-axis \cite{lab44}.

%---------
\subsection{ Performance evaluation}

Fidelity is an important measure, which is used to evaluate the performance of the protection schemes. Fidelity between the initial state $\rho _{in} $  and the final state of the system after the whole protection scheme$\rho _{f} $ is given as:
\begin{equation}\label{eq14} 
Fid=Tr\sqrt{\sqrt{\rho _{in} } \rho _{f} \sqrt{\rho _{in} } }  
\end{equation}

Fidelity $Fid\in [0,1]$ measures how much two states overlap each other. A fidelity of 1 means the states are identical, whereas the fidelity of 0 means the states are orthogonal. The aim of protection schemes is to bring the final state as close as possible to the initial state, which means gaining complete fidelity close to 1. 

In case of entanglement protection, the degree of entanglement is described by the so-called concurrence to quantify the effect of protection \cite{lab45}. Concurrence of the state $\rho $ is defined by
\begin{equation}\label{eq15} 
C\left(\rho \right)=max\left\{0,{\lambda }_1-{\lambda }_2-{\lambda }_3-{\lambda }_4\right\}
\end{equation}
where $\lambda _{i} $'s are, in decreasing order, the nonnegative square roots of the eigenvalues of the matrix $\rho \tilde{\rho }$. Here $\tilde{\rho }$ is the matrix given by
\begin{equation}\label{eq16} 
\tilde{\rho }\equiv \left(\sigma _{y} \otimes \sigma _{y} \right)\rho ^{*} \left(\sigma _{y} \otimes \sigma _{y} \right)
\end{equation}
where $\rho ^{*}$ denotes the complex conjugate.

\section{ Weak measurement and quantum measurement reversal}\label{sec3}

\begin{figure}[h]
\centering
  \includegraphics[width=\textwidth]{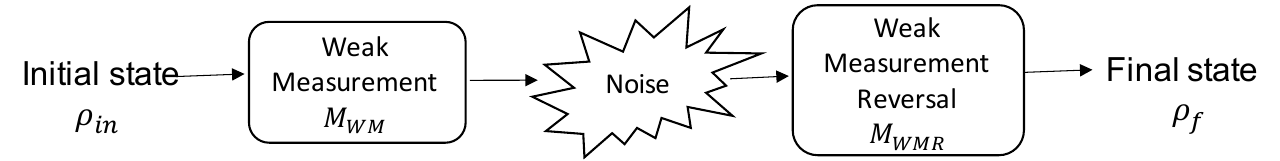}
\caption{schematic diagram of WMQMR scheme.}
\label{fig1}      
\end{figure}

The notion of combining WM and QMR is first introduced in \cite{lab23}, where the WM must have adjusting strength to make the system less disturbed. In other words, by applying WM the information extracted from a quantum system is deliberately limited, to avoid the measured system's state from randomly collapsing towards an eigenstate. Therefore it would be possible to restore the initial state by applying some operations \cite{lab46,lab46a,lab46b,lab46c,lab46d,lab46e}. In WMQMR protection scheme, a WM and its reversing measurement are introduced before and after the decoherence channel, respectively \cite{lab23}. The schematic diagram of WMQMR scheme is represented in Fig. \ref{fig1}.

In this scheme, first, a WM is applied on the qubit to move it toward the ground state and reduce the weight of the excited state $\lvert 1\rangle$ (to make it less vulnerable to the effects of the noise). The corresponding map of the WM with strength $p_1$ on the qubit in the computational basis $\left\{\lvert 0\rangle, \lvert 1\rangle\right\}$ can be written as

\begin{equation}\label{eq17} 
M_{WM}=\left( \begin{array}{cc}
1 & 0 \\ 
0 & \sqrt{1-p_1} \end{array}
\right)
\end{equation}

After the decoherence the QMR is applied to restore the weight of the excited state and recover the initial state. The corresponding map of the QMR with strength $p_2$ in the computational basis can be represented as

\begin{equation}\label{eq18} 
M_{QMR}=\left( \begin{array}{cc}
\sqrt{1-p_2} & 0 \\ 
0 & 1 \end{array}
\right)
\end{equation}

To perform the WM, a device is needed to monitor the qubit indirectly. When the device has no signal (called as null result case), the qubit is partially collapsed and the WM has successfully applied. However, if the device has signals the result must be discarded \cite{lab47}. Due to the failure rate of the WM the WMQMR is a probabilistic scheme with limited success probability. Hence in WMQMR the tradeoff between success probability and fidelity is unavoidable which depends on the strength of the WM. In fact, there is a tradeoff between information gain and reversibility in WM which is proved in \cite{lab48,lab48a,lab48b} and is one of the distinct features of WM. Particularly, the full trade-off relations among information gain, state disturbance and reversibility of WMQMR is driven in \cite{lab48c}.

\begin{figure}[h]
\centering
  \includegraphics[width=0.7\textwidth]{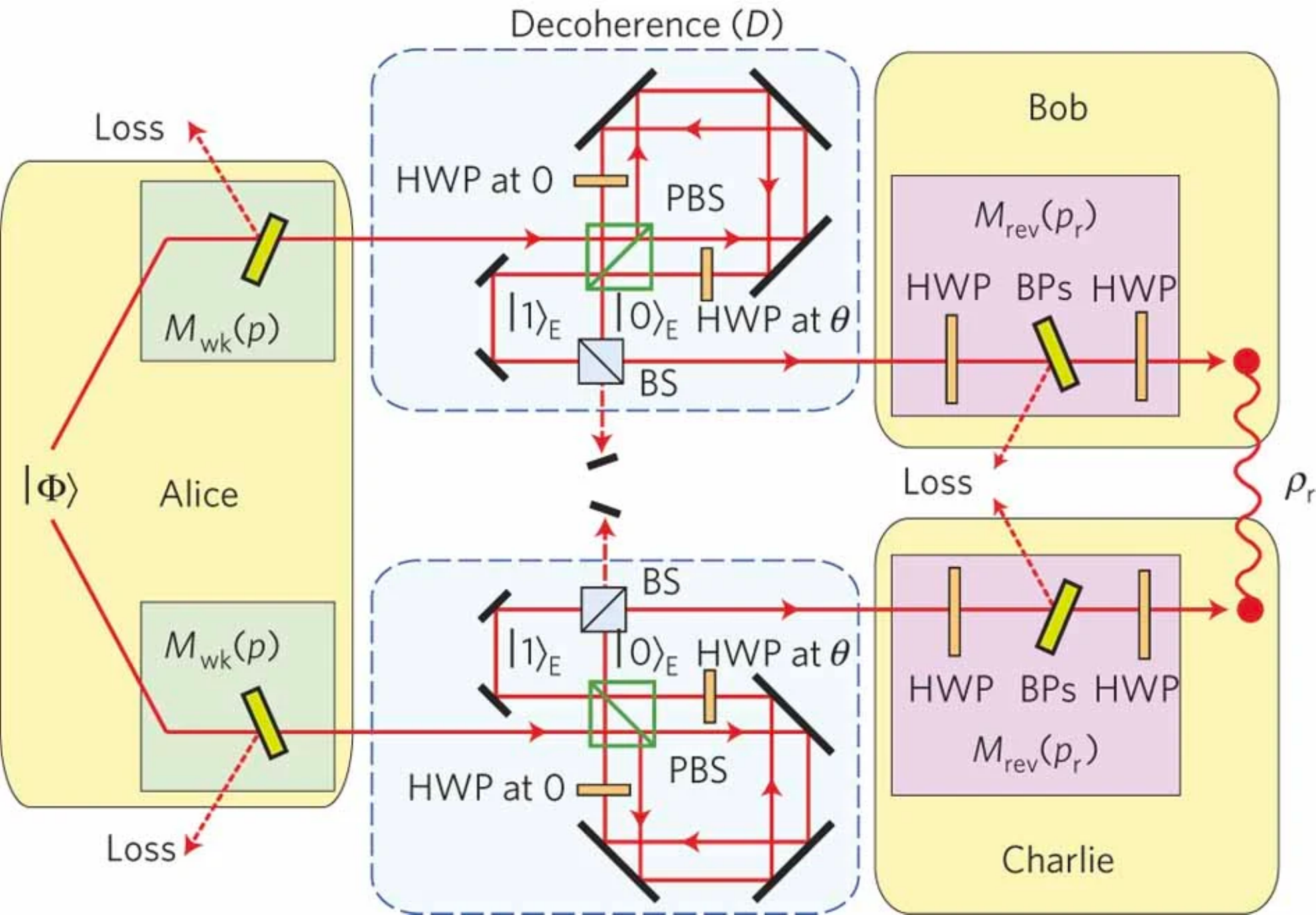}
\caption{WMQMR experiment diagram \cite{lab11}.}
\label{fig2}      
\end{figure}

The WMQMR has been experimentally implemented for one photonic qubit in \cite{lab49,lab50} as well as superconducting phase qubits \cite{lab22,lab51} and trapped-ion \cite{lab51a}. Moreover, to protect the entanglement as a vital resource in quantum systems, the WMQMR is investigated both theoretically \cite{lab52,lab53,lab54,lab55} and experimentally \cite{lab11}. In \cite{lab56} it is assumed that both half of the entangled pair undergo separate decoherence, which decreases the amount of the entanglement and often causes ESD. It is shown that it's enough to apply the WMQMR on one qubit of the entangled pair instead of both qubits to avoid ESD. In \cite{lab11},  the experimental implementation of entanglement protection from amplitude-damping decoherence via WMQMR is given. The experimental diagram is demonstrated in Fig. \ref{fig2} where the initial two-qubit is prepared with the two-photon polarization state. The WM and QMR are performed using Brewster-angle glass plates (BPs) and half-wave plates (HWPs); And the AD is implemented using an interferometer.

It is shown that the scheme can compensate the effects of decoherenece by increasing the amount of concurrence. Also, the trade-off relation between the success probability, concurrence and WM strength has been studied. For larger WM strength the amount of concurrence is higher while it gives less success probability.

The WMQMR has been widely applied to various aspects of quantum information processing. It can effectively protect quantum entanglement of the single-qubit \cite{lab56a}, two qubit \cite{lab46f,lab56b},three-qubit \cite{lab56c,lab56d} and high dimensional systems \cite{lab56e} from different damping channels. Moreover, the WMQMR has been used in quantum teleportation to protect the entangled shared state in damping channels \cite{lab56f,lab56g,lab56h,lab56i}. It has been shown that WMQMR can enhance the teleportation fidelity significantly in presence of different noise types and correlated decoherence channels compared to teleportation without any protection \cite{lab56j,lab56l,lab56m,lab56n,lab56o,lab56p}.

\section{ Quantum feedback control}\label{sec4}

The idea of using quantum WM and feedback control to protect the quantum state in the PD channel is put forward in \cite{lab27}. The initial state is one of two non-orthogonal states in the $x$-$z$ plane of the Bloch sphere as:
\begin{equation}\label{eq19} 
{\lvert \psi _{\pm }\rangle} =\cos \left(\alpha /2\right){\lvert+\rangle} \pm \sin \left(\alpha /2\right){\lvert - \rangle} 
\end{equation}

The PD noise is considered as "kicking'' the state of the qubit in one direction or the other on the Bloch sphere as given in Eq. \eqref{eq2}. The state goes through the PD channel and then is measured by measurement operators along the $y$-axis $M_{y} $ as given in Eq. \eqref{eq12}, to gain information about the direction of the noise kick. By using quantum WM they balance a tradeoff between information gain and disturbance caused by the measurement. In the last step, according to the result of the measurement, the correction rotation is applied about the $z$-axis which is $R_{z} $ in Eq. \eqref{eq13}. The whole control procedure is given by a quantum operation as a completely positive trace-preserving (CPTP) map:

\begin{equation}\label{eq20} 
C(\rho _{e} )=\left(R_{z+} M_{y+} \right)\rho _{e} \left(R_{z+} M_{y+} \right)^{\dag } +\left(R_{z-} M_{y-} \right)\rho _{e} \left(R_{z-} M_{y-} \right)^{\dag } 
\end{equation}
where $\rho _{e} $ is the damped state of the system given in Eq. \eqref{eq3}. 

\begin{figure}[h]
\centering
  \includegraphics[width=0.8\textwidth]{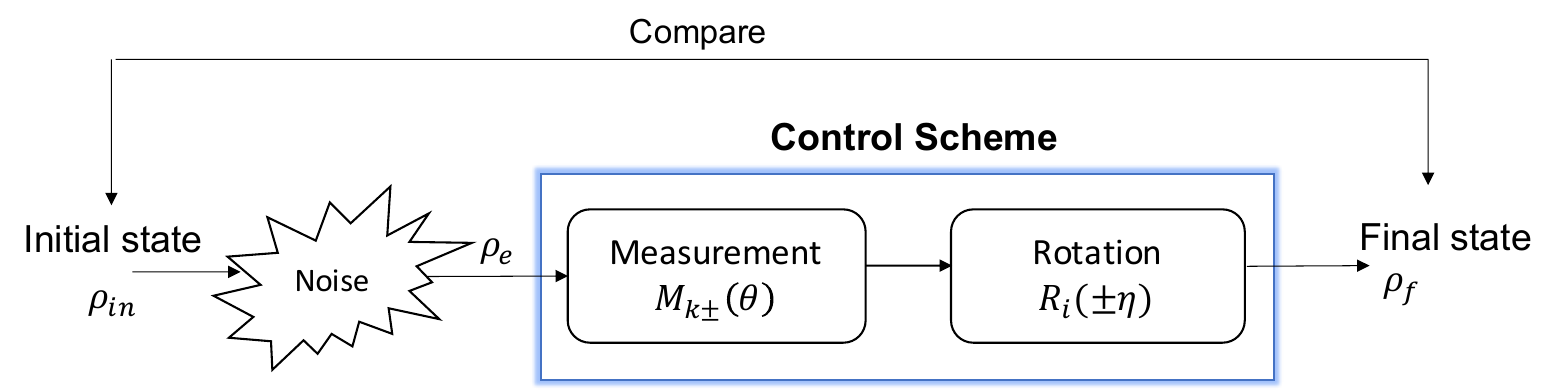}
\caption{The schematic diagram of QFBC.}
\label{fig3}      
\end{figure}

The performance of the scheme is evaluated by fidelity between initial state and the final state after the whole process of protection as defined in Eq. \eqref{eq14}. The fidelity can be optimized over rotation angle to find the optimum rotation angle $\eta_{opt}\left(\alpha,r, \theta\right)$ as a function of initial states angle $\alpha $, damping probability $r$ and measurement angle $\theta $. Moreover, one can find the optimum measurement strength to achieve the best trade-off between gaining information about the system and disturbing it through measurement back-action by substituting the optimum rotation angle in fidelity equation. The optimum measurement angle $\theta _{opt} (\alpha ,r)$ is a function of initial states angle $\alpha $ and damping probability $r$. Hence, to apply the optimum QFBC, the initial state and the noise channel must be fully characterized and completely known. The schematic diagram of QFBC is given in Fig. \ref{fig3}. The term "feedback" refers to a subsequent operation (rotation) performed on the same state (the damped state) as shown in Eq. \eqref{eq20}, which is standard in the quantum control literature.

\begin{figure}[h]
\centering
  \includegraphics[width=0.7\textwidth]{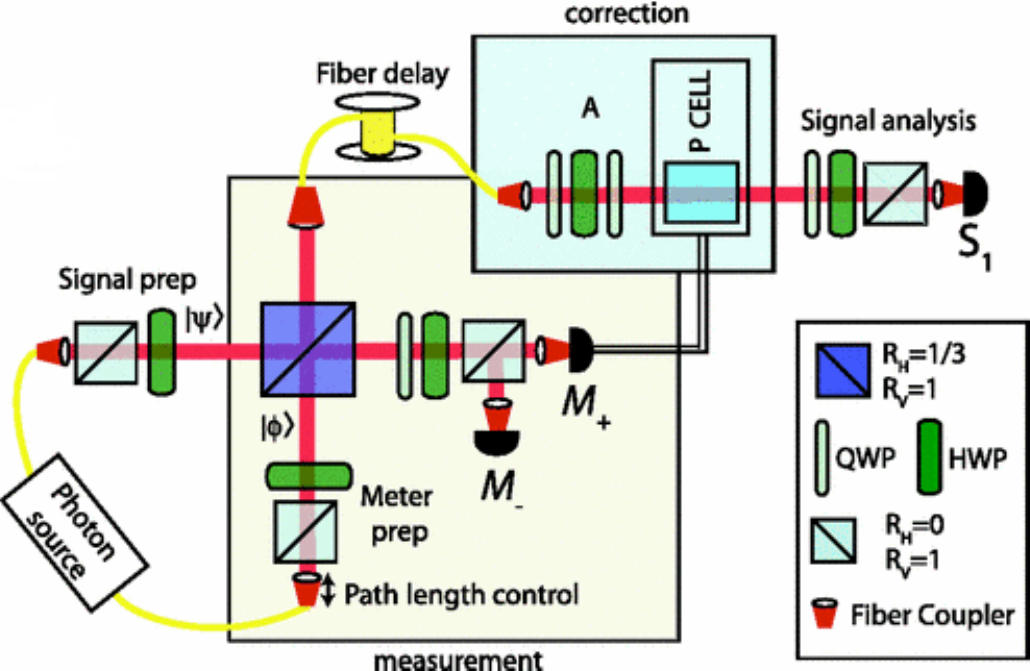}
\caption{QFBC experimental diagram \cite{lab26}.}
\label{fig4}      
\end{figure}

The experimental realization of the proposed QFBC is given in \cite{lab26}. Again, the initial state is one of two non-orthogonal states in the $x$-$z$ plane, but they use measurement operator along the $z$-axis $M_{z_{\pm } } $ and rotation along the $y$-axis $R_{y} (\pm \eta )$. Fig. \ref{fig4} depicts the experimental diagram where a signal qubit which is encoded in the polarization of a single photon, passes through a phase damping channel. The required variable-strength measurement (including strong and WM) on the signal qubit is realized by entangling it to another meter qubit (photon) using a nondeterministic linear optic controlled-Z (CZ) gate, and then a full strength projective measurement on the meter qubit is performed. This implements a measurement on the signal qubit with a strength determined by the input meter state. Finally, the outcome of the projective measurement determines the correction rotation on the signal qubit.

To study protecting more general initial states, one can define the initial states as:
\begin{equation}\label{eq21} 
{\lvert \psi \rangle} =\cos \left(\alpha /2\right){\lvert + \rangle} +e^{i\phi } \sin \left(\alpha /2\right){\lvert - \rangle}  
\end{equation}

By considering the angle $\phi $, the initial states can lie in any plane of the Bloch sphere. In \cite{lab57}, protection of these more general initial states, in the presence of different noise sources such as bit-flip noise, AD noise, PD noise, depolarizing noise is given. The measurement operators of QFBC are along the $z$-axis and the rotation along the $y$-axis is applied to recover the state. The comparison between QFBC via WM, do nothing, and projective measurement shows that for some suitable states, the QFBC using WM could outperform the do nothing and projective measurement schemes for all typical types of noise sources.

Furthermore, the question arose if the WM used in previous schemes is also the best one for the protection of general states. Hence an extended scheme is given in \cite{lab58}, where the initial states are two non-orthogonal states 
\begin{equation}\label{eq22} 
{\lvert \psi _{\pm }  \rangle} =\cos \left(\alpha /2\right){\lvert + \rangle} \pm e^{i\phi } \sin \left(\alpha /2\right){\lvert - \rangle} 
\end{equation}

 They generalize the scheme by using the new set of measurements to the feedback control:
\begin{equation}\label{eq23} 
\begin{array}{l} {M_{+} \left(\theta \right)=\cos \left(\theta /2\right){\lvert 0 \rangle} {\langle 0 \rvert} +e^{i\beta } \sin \left(\theta /2\right){\lvert 1 \rangle} {\langle 1 \rvert} } \\ {M_{-} \left(\theta \right)=e^{i\beta } \sin \left(\theta /2\right){\lvert 0 \rangle} {\langle 0 \rvert} +\cos \left(\theta /2\right){\lvert 1 \rangle} {\langle 1 \rvert} } \end{array} 
\end{equation}

The parameter $\beta $ allows mixing the $x$- and $y$- components of the Bloch sphere that improves the performance of the state protection for protecting more general states in Eq. \eqref{eq22}. It is shown that measurement operators without parameter $\beta $ are not the best for protection of general states. The maximum performance of the QFBC with generalized measurements are for initial states with parameters around $\alpha ={\pi \mathord{\left/ {\vphantom {\pi  4}} \right. \kern-\nulldelimiterspace} 4} $ and $\phi ={\pi \mathord{\left/ {\vphantom {\pi  4}} \right. \kern-\nulldelimiterspace} 4} $ in Eq. \eqref{eq22}.

In all the above approaches, Kraus operators are used to stand for noise and the feedback control is described by a control map. In \cite{lab59,lab60}, the QFBC is considered for an open quantum system where a master equation is derived in the Lindblad form describing the evolution of the open quantum system subjected to a QFBC. The initial states are known mixed states composed of two nonorthogonal states. They found that by using proper QFBC scheme any initial pure or mixed state can effectively drive into an arbitrary given target pure state. For the initial state prepared in one of two nonorthogonal states, the optimal feedback with WM is more effective to protect the system against decoherence than the one with projective measurement and do-nothing, which is consistent with the results of Ref \cite{lab26,lab27}. They have shown that if the initial state is known, no matter mixed or pure, the QFBC scheme based on projective measurement is much better, and if the protected state is not completely known, the QFBC scheme based on WM is more effective. 

In most previous schemes, the QFBC is considered in $y$- and $z$- basis of the Bloch sphere. While for arbitrary initial states, which can be in any plane of the Bloch sphere one needs to consider all bases of the Bloch sphere during the control procedure.  In \cite{lab38}, the optimum control parameter for measurement and rotation along different bases of the Bloch sphere are given. They find that depending on the location of the initial state on the Bloch sphere one needs to apply control in different bases to get the best protection.

 Moreover, the QFBC is combined with other protection methods such as MED to protect two nonorthogonal states against decoherence \cite{lab62}.

\section{ Quantum feed forward control}\label{sec5}

In this section, we review the noise protection schemes based on feedforward control. In most feedback protection schemes, the initial state is known and the optimum parameters of the control are dependent on the initial state variables. It has been proved that prior information about the state of the system is essential for optimum protection \cite{lab63,lab64,lab65}. In QFFC the focus is not on the initial state, but on the damping channel. In QFBC the intention of measurement is to acquire information about the initial state while in QFFC the measurement is done to transfer the state to a desired target state which is less vulnerable to the effect of the damping channel followed by flip operation. In fact, the WMQMR and QFFC are following the same strategy, while in QFFC by adding the flip operations after the WM, the state is certainly transferred to a less vulnerable state.

The first idea of QFFC is given by Wang et al. to protect arbitrary unknown states from AD \cite{lab24}. Before the state goes through the noise channel, they apply WMs to get some information about the initial state. The measurement operators in this scheme are as follows:

\begin{equation}\label{eq24} 
M_{1} =\left(\begin{array}{cc} {\sqrt{p} } & {0} \\ {0} & {\sqrt{1-p} } \end{array}\right),\, M_{2} =\left(\begin{array}{cc} {\sqrt{1-p} } & {0} \\ {0} & {\sqrt{p} } \end{array}\right) 
\end{equation}
where $p$ is the strength of the measurement. These measurement operators are along the $z$-axis and equal to measurement operators $M_{z\pm } $ in Eq. \eqref{eq12} with $p=\cos ^{2} \left({\theta \mathord{\left/ {\vphantom {\theta  2}} \right. \kern-\nulldelimiterspace} 2} \right)$. 

Then according to the result of the measurement, the feed forward control is applied to transfer the system to a target state, which is almost immune to the effect of the noise. The feed forward operators are as follow:
\begin{equation}\label{eq25} 
F_{1} =\left(\begin{array}{cc} {1} & {0} \\ {0} & {1} \end{array}\right),\, \, F_{2} =\left(\begin{array}{cc} {0} & {1} \\ {1} & {0} \end{array}\right) 
\end{equation}
where $F_{1} \, ( \text{or }F_{2} )$ corresponds to the result of measurement operator $M_{1} \, (\text{or }M_{2} )$. If the result corresponds to $M_{1} $is acquired, it means the state is already close to the ground state and noise does not have much effect on the state. Hence, the feedforward operation $F_{1} $ is chosen to be the identity operator. However, when the result corresponding to the measurement operator $M_{2} $is acquired, by applying the feed forward operator $F_{2} $,  the state of the system gets closer to the ground state and robust to the effects of the AD channel.

After applying the measurement and feedforward operation, the state goes through the AD channel. Then, to recover the state as close as possible to the initial state, the reversed feed forward operators are applied which are the same as $F_{1} $ and $F_{2} $ given in Eq. \eqref{eq25}. 

\begin{figure}[h]
\centering
  \includegraphics[width=\textwidth]{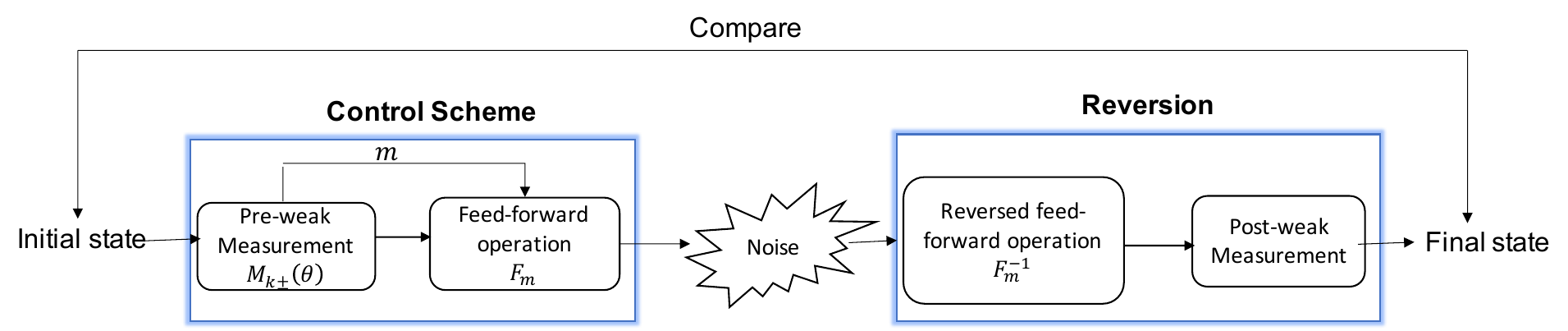}
%\captionsetup{justification=centering}  
\caption{Schematic diagram of QFFC.}
\label{fig5}      
\end{figure}

In the last step of protection, the post-WM is applied using partial WM, $\Lambda =N_{1}^{\dag } N_{1} $ and $\Xi =W_{1}^{\dag } W_{1} $, where $N_{1} $ and $W_{1} $ are defined as:
\begin{equation}\label{eq26} 
N_{1} =\left(\begin{array}{cc} {\sqrt{1-p_{u} } } & {0} \\ {0} & {1} \end{array}\right),\, \, W_{1} =\left(\begin{array}{cc} {1} & {0} \\ {0} & {\sqrt{1-p_{v} } } \end{array}\right) 
\end{equation}
where $N_{1} \, ( \text{or }W_{1} )$ corresponds to the result of measurement operator $M_{1} \, ( \text{or }M_{2} )$ respectively; and  $p_{u} ,\, p_{v} $are the post-WM strength. The post-WMs are approximately the reversal of the corresponding pre-WM, $M_1N_1\sim I$ and $M_2W_1\sim I$. The schematic diagram of QFFC is given in Fig. \ref{fig5}.

 Because of the partial WMs in the last step of the protection scheme, QFFC becomes a probabilistic scheme with success probability. It can protect two arbitrary nonorthogonal states with an almost unit fidelity but with the price of low success probability. 

The QFFC is extended to protect the arbitrary state of a two-qubit system \cite{lab66}. It is shown that the WM scheme could improve the protection for both pure and mixed initial states even for intense damping probabilities. In addition, the complete recovery condition and total success probability are independent of initial state, while total fidelity depends on the initial state variables. The comparison between QFFC and a protection scheme based on quantum gates in \cite{lab67} shows that the QFFC always has significant improvement in terms of success probability (fidelity) for the given fidelity (success probability).

In \cite{lab68} the QFFC is used to realize a better effect of discrimination for any pair of nonorthogonal states and any degree of amplitude damping. It is shown that the QFFC discrimination is advantageous over the minimum-error (ME) discrimination for any two initial nonorthogonal states, any degree of AD noise and any prior probabilities. In fact, the heavier the AD is, the more superior the QFFC scheme is.

As we mentioned before QFFC can achieve high fidelity at the price of low success probability. Therefore, the practical question is: whether there is a quantum state protection scheme that can obtain high fidelity while avoiding the cost of success probability. To answer this question, in \cite{lab69} the QFBC and QFFC are combined to protect two nonorthogonal states in AD channel. In the last step of QFFC, according to the result of the post-WM, the feedback control is added as a rotation operation along the $y$-axis to recover the state. In fact, the post-WMs in QFFC process are substituted for the measurement in the QFBC process. The comparison between composite control and previous protection schemes (QFBC and QFFC), shows that the composite control can achieve the biggest fidelity for given success probability, and for given fidelity, it has the biggest success probability. 

Later in \cite{lab70} the authors use the same scheme as QFFC but without post-WM to protect quantum Fisher information of N-qubit GHz state. Before the AD channel, the same measurements and operators as QFFC are applied but after the AD channel, they only use the reversed operators to recover the state. They call the scheme as WM with pre- and post-flips (WMPPF). Since there are no partial measurements in the protection scheme, the WMPPF always has the probability of 1. This scheme is compared with WM and quantum measurement reversal (WMQMR) protection scheme. The comparison result shows that the WMPPF can reduce the effect of dissipation on the average QFI of the phase or the frequency for GHZ state and some generalized GHZ states, and the WMQMR can reduce the effect of dissipation on the average fidelity for GHZ state and generalized GHZ states in AD channel. Comparing QFI with fidelity for WMPPF or for WMQMR, a scheme protecting the average fidelity does not necessarily protect the average QFI, even with the same parameters, and vice versa.

In \cite{lab25}, instead of using post-partial measurements the rotation operators are used to protect the unknown $N$-qubit state. This scheme can highly improve the effect of state protection compared with the earlier scheme in \cite{lab24} and is extended to protect the state of N-qubit system. It has been shown that by increasing the number of qubits, the behavior of total fidelity improves. 

The QFFC can also be used to protect the nonlocality of the Bell and GHZ states against dissipation \cite{lab71}. Quantum nonlocality refers to the property that the correlation of measurement results between two distant systems violates the local hidden-variable (LHV) theory \cite{lab72,lab73}. Quantum nonlocality in two-partite case is usually characterized by the Bell inequality, which is used in a wide range of protocols for quantum information processing \cite{lab74,lab75,lab76,lab77}. The QFFC protection scheme can increase the noise threshold from 0.5 to 0.98 for Bell state, and for GHZ state from 0.29 to 0.96. In addition, it can protect entanglement relatively easier than nonlocality.

\section{ Comparison between QFBC and QFFC }\label{sec6}

The analytical comparison between QFBC and QFFC for protection of the state against different types of noise sources: AD noise, PD noise, bit-flip noise, and depolarizing noise, has been done in Ref  \cite{lab64,lab65,lab78}. It has been mathematically proved that without knowledge about the initial state, one can not suppress the noise effectively with only QFBC and feedforward operations are essential to protect the unknown initial state. 

As it has been shown in \cite{lab24}, the QFFC outperforms the WMQMR even for intense decaying rates. Hence, in this section we only consider the comparison between QFFC and QFBC. We compare the performance of QFBC and QFFC for protection against both AD and PD by numerical simulations. Here we assume that the initial state and damping channel are completely known. We find the optimal QFBC and QFFC to protect completely known initial states through characterized PD and AD noise channels.

To quantify the difference of performance between QFBC and QFFC, we define the measure of difference as
\begin{equation} \label{eq27} 
F_{diff} =F_{QFBC}^{opt} -F_{QFFC}^{opt}  
\end{equation} 
where $F_{QFBC}^{opt} $ is the optimum fidelity of the QFBC scheme and $F_{QFFC}^{opt} $ is the optimum fidelity of QFFC protection scheme. In fact, we apply the QFBC and QFFC for protection of the known state from both AD and PD separately. Then we find the difference of their optimal performances as Eq. \eqref{eq27} to compare them with each other. Here, QFBC refers to the protection scheme in \cite{lab44} where both AD and PD are considered; and QFFC denotes the scheme in \cite{lab25}.

\begin{figure*}[ht!]
    \centering
    \begin{minipage}[t]{0.3\textwidth}
        \centerline{\includegraphics[width=\textwidth]{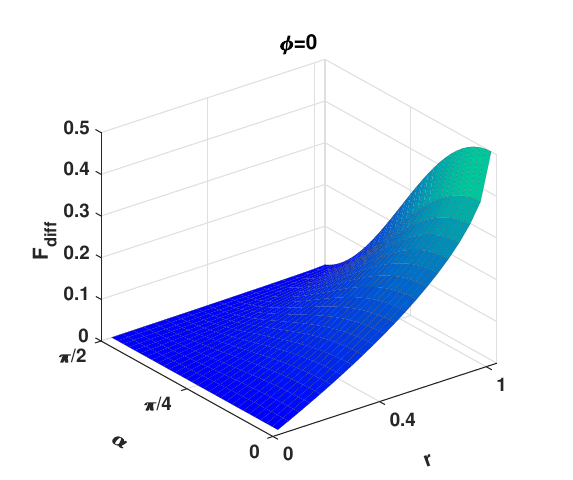}}
        \centerline{(a)}
    \end{minipage}
    \hfill
    \begin{minipage}[t]{0.3\textwidth}
        \centering{\includegraphics[width=\textwidth]{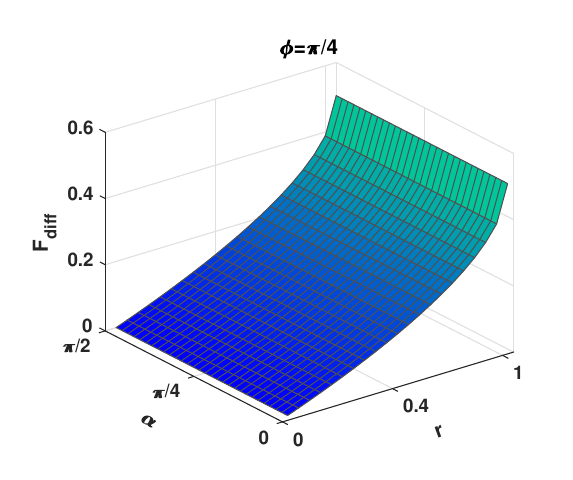}}
        \centering{(b)}
    \end{minipage}
    \hfill
    \begin{minipage}[t]{0.3\textwidth}
        \centering{\includegraphics[width=\textwidth]{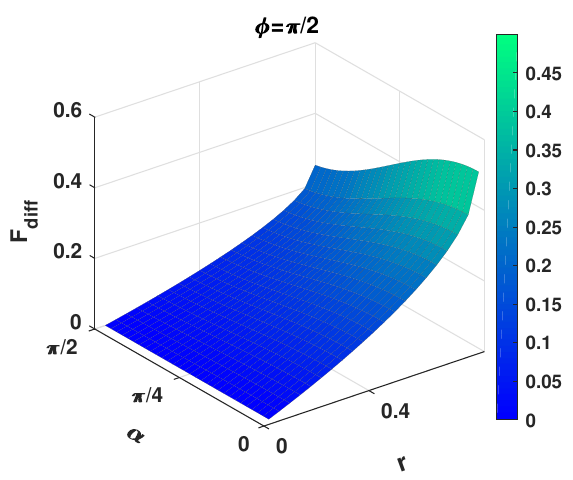}}
        \centering{(c)}
    \end{minipage}
    \hfill
    \begin{minipage}[t]{0.3\textwidth}
        \centering{\includegraphics[width=\textwidth]{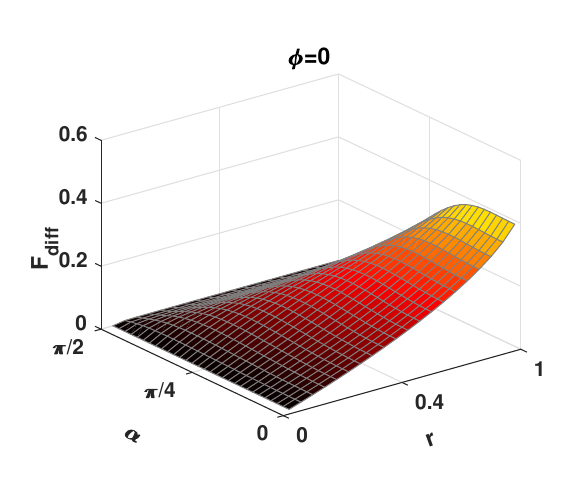}}
        \centering{(d)}
    \end{minipage}
    \hfill
    \begin{minipage}[t]{0.3\textwidth}
        \centering{\includegraphics[width=\textwidth]{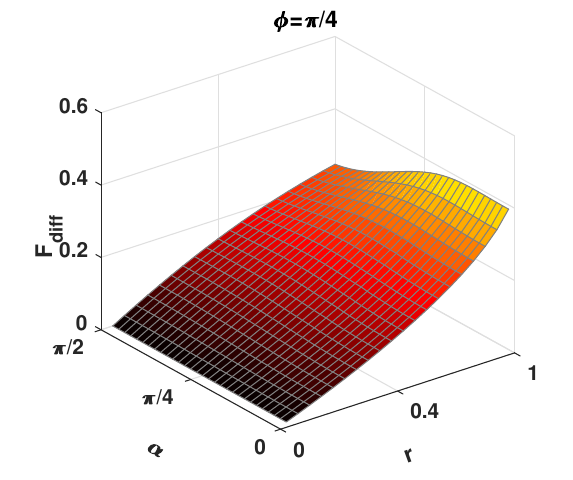}}
        \centering{(e)}
    \end{minipage}
    \hfill
    \begin{minipage}[t]{0.3\textwidth}
        \centering{\includegraphics[width=\textwidth]{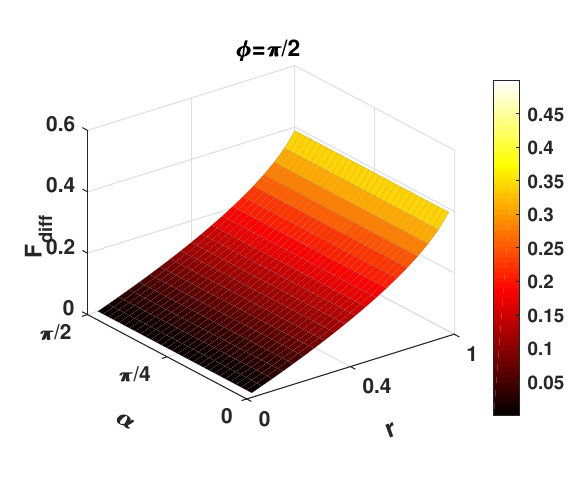}}
        \centering{(f)}
    \end{minipage}
\caption{$F_{diff}$ as a function of initial state angel $\alpha$ and damping probability $r$  (PD and AD noise) for fixed amount of $\phi=0, \frac{\pi}{4}, \frac{\pi}{2}$. The blue planes indicate the difference of the optimal performance of QFBC and QFFC for protection against PD and red planes are the difference of the optimal performance of QFBC and QFFC for protection against AD. }\label{fig6}
\end{figure*}

We perform the numerical simulations for $F_{diff} $ by changing the initial state angle $\alpha $ from 0 to ${\pi \mathord{\left/ {\vphantom {\pi  2}} \right. \kern-\nulldelimiterspace} 2} $ in Eq. \eqref{eq21} and also damping probability $p$ uniformly varies between 0 and 1 in Eqs. \eqref{eq4} and \eqref{eq10} identify PD and AD noise, respectively. To find the optimum fidelity for each amount of initial state angle and damping probability we grid the range of measurement angle and rotation angle between 0 and ${\pi \mathord{\left/ {\vphantom {\pi  2}} \right. \kern-\nulldelimiterspace} 2} $ with steps of ${\pi \mathord{\left/ {\vphantom {\pi  60}} \right. \kern-\nulldelimiterspace} 60} $. We emphasize that we do not use the optimum mathematical formula for control parameters which is given in different schemes, but we find the optimum fidelity by considering all possible control parameters. The results for fixed amount of $\phi $=0, ${\pi \mathord{\left/ {\vphantom {\pi  4}} \right. \kern-\nulldelimiterspace} 4} \, \, {\rm and}{\, \pi \mathord{\left/ {\vphantom {\, \pi  2}} \right. \kern-\nulldelimiterspace} 2} $ are given in Fig. \ref{fig6}.

As Fig. \ref{fig6} demonstrates, $F_{diff} $ always has positive values, which means that QFBC performs better than QFFC for all states and damping probabilities for protection against both AD and PD. For initial states with $\phi =0$, the difference of optimum fidelities becomes smaller when $\alpha $ gets closer to ${\raise0.7ex\hbox{$ \pi  $}\!\mathord{\left/ {\vphantom {\pi  2}} \right. \kern-\nulldelimiterspace}\!\lower0.7ex\hbox{$ 2 $}} $. For $\phi ={\pi \mathord{\left/ {\vphantom {\pi  4}} \right. \kern-\nulldelimiterspace} 4} $and ${\pi \mathord{\left/ {\vphantom {\pi  2}} \right. \kern-\nulldelimiterspace} 2} $,  by increasing the damping probability the amount of $F_{diff} $ increases to 0.5. Hence the QFBC has almost 50\% better performance for higher damping probabilities than QFFC. Once more, we note that here we assume the initial state and damping channel are completely known. In addition, to find the optimum fidelity we did not use the optimum formula of control parameters given in different schemes, but we find it by numerical simulations over all possible measurement and rotation angles for each damping probability and initial state.  

\section{Conclusion}\label{sec7}

 We demonstrated different definitions of the two most important types of noise sources: AD and PD, to better understand the effect of noise on the state of the system. PD noise makes the state rotate in one or other direction in Bloch Sphere representation. Hence, in the control procedure, first one needs to gain information about the direction of the rotation and then apply rotation to bring the state back to its initial one. While AD noise has less effects on states close to ground of the system; therefore, by applying WM and flip operations before the noise channel, the state goes closer to the ground state to become less vulnerable and after the noise channel the reversed operations are applied to recover the state. We reviewed three different measurement base protection schemes: WMQMR, QFBC and QFFC. In WMQMR a WM and QMR are applied before and after decoherence, respectively. While QFFC adds flip operations to make the initial state less vulnerable to the effects of the noise. The behind physics is to project the initial state to a more robust state by applying WM in WMQMR and WM+flip operations in QFFC. However, QFBC follows a different strategy, it measures the state after the decoherenece and applies the correction operations to recover the initial state.

We compared the QFBC and QFFC in the case that the initial states and damping channels are completely characterized and known. The comparison results demonstrate that QFBC outperforms the QFFC if the initial state and damping channel are completely known.

%---------------------------------------
\backmatter

\bmhead{Acknowledgments}
This work was partially supported by the NationalNatural Science Foundation of China under Grant  61973290 and Grant 61720106009. The research of J. J. Nieto has been partially supported by the Agencia Estatal de Investigacion (AEI) of Spain, project PID2020-113275GB-I00, co-financed by the European Fund for Regional Development (FEDER); and by Xunta de Galicia under grant ED431C 2019/02.

\end{document}